\documentclass[aps,pra,reprint,superscriptaddress]{revtex4-2}

\usepackage[colorlinks,breaklinks,linkcolor=blue,anchorcolor=blue,citecolor=blue,urlcolor=blue]{hyperref}
\usepackage{graphicx}
\usepackage{amsmath}

\begin{document}


\title{Unconventional Photon Blockade in Second-Order Nonlinearity Coupled Cavities with Asymmetric Dissipation}


\author{Zhi-Qiang Zhang}
\email[Corresponding author. E-mail: ]{zhangzhiqiang08@gmail.com}
\affiliation{Department of College Physics, School of Intelligent Construction, Zhengzhou Business University, Zhengzhou 451200, China}

\author{Xiao-Ai Zhang}
\affiliation{School of Computer and Big Data Engineering, Zhengzhou Business University, Zhengzhou 451200, China}

\author{Yan-Zhe Song}
\affiliation{Beijing Hangyu Communication Co., LTD. Jilin Branch Changchun 130000, China}

\author{Meng-Li Guo}
\affiliation{School of Science, East China University of Technology, Nanchang 330006, China}

\author{Yan-Hui Zhou}
\email[Corresponding author. E-mail: ]{yanhuizhou@126.com}
\affiliation{Quantum Information Research Center and Jiangxi Province Key Laboratory of Applied Optical Technology, Shangrao Normal University, Shangrao 334001, China}

\date{\today}

\begin{abstract}
We investigate the unconventional photon blockade in a 
second-order nonlinearity coupled double-cavity system with asymmetric 
dissipation. By treating the cavity decay rates as independent parameters, 
we analytically derive two exact families of optimal blockade conditions. 
While the first family recovers and generalizes the symmetric-dissipation 
results of previous studies, the second family, arising uniquely from an 
asymmetric dissipation ratio of \(\kappa_B = 2\kappa_A\), completely locks 
the relative phase of the interference pathways. Importantly, we find that 
this asymmetric dissipation not only renders the optimal blockade condition 
strictly immune to frequency detuning, but also enhances the blockade depth 
by an order of magnitude compared to the resonant case. Furthermore, the 
mean photon number remains almost unaffected, proving that 
this enhanced robustness is achieved without sacrificing the brightness 
of the single-photon source. These findings offer a promising route toward 
robust single-photon generation in passive cavities with second-order nonlinearity.
\end{abstract}


\maketitle

\section{Introduction}
Efficient and deterministic single-photon sources are indispensable 
resources for a wide range of applications in quantum information 
science, including linear optical quantum computing \cite{Kok2007}, quantum 
cryptography \cite{Gisin2002}, and quantum metrology \cite{Couteau2023,Holewa2025}. A fundamental mechanism 
for generating such nonclassical light is the photon blockade (PB) 
effect, wherein the presence of a single photon inside a nonlinear 
cavity shifts the resonance frequency, thereby preventing the entry 
of a second photon \cite{Carusotto2013,PhysRevLett.79.1467,Barak2008,Birnbaum2005}. In the conventional photon blockade 
(CPB) regime, this anharmonicity originates from strong single-photon 
nonlinearities, which induce a significant energy-level splitting that 
far exceeds the cavity linewidth \cite{Verger2006,Lang2011,Fink2009}. Theoretically, this is 
often modeled by strong light-matter interactions in cavity quantum 
electrodynamics (cQED) or by strong Kerr-type nonlinearities, where 
the system behaves as a two-level emitter to the incoming photons. 
Despite its conceptual simplicity and high fidelity, achieving strong 
enough optical nonlinearities at the single-photon level remains a 
formidable experimental challenge. This stringent requirement severely 
limits the practical scalability of CPB-based devices, as it necessitates 
either ultra-high-\(Q\) factors, extremely small mode volumes, 
or exotic materials that often suffer from significant inherent losses.

To circumvent the stringent requirement for strong nonlinearities, 
an alternative mechanism known as unconventional photon blockade (UPB) 
has been proposed~\cite{Liew2010}. Unlike CPB, which relies on anharmonic 
energy-level splitting, UPB emerges from destructive quantum 
interference between multiple distinct transition pathways to the 
two-photon state \cite{Flayac2017,Bamba2011}. Specifically, when the system is weakly 
driven, the two-photon state can be reached through multiple competing 
pathways, typically including the direct two-photon excitation of the 
cavity mode, cascaded transitions mediated by intermediate states, and 
other higher-order scattering processes. 
The total transition amplitude to the two-photon state is governed by 
the coherent superposition of these individual pathways. By appropriately 
engineering the system such that the relative phase between these pathways 
is strictly \(\pi\), the amplitudes undergo perfect destructive interference, 
thereby preventing the two-photon excitation even in the weak-nonlinearity 
regime. This multipath interference mechanism circumvents the stringent 
prerequisite for strong nonlinearities, thereby facilitating the generation 
of nonclassical light in weakly nonlinear platforms.

Over the past decade, the UPB has been 
extensively investigated in both theoretical and experimental studies 
across diverse physical platforms. For instance, Snijders 
\textit{et al.} observed the UPB in a quantum dot cQED system by 
utilizing a single quantum dot transition coupled to two orthogonally 
polarized cavity modes \cite{Snijders2018}, while Vaneph \textit{et al.} 
realized this effect in the microwave domain using two coupled 
superconducting resonators \cite{Vaneph2018}. In addition to the quantum 
interference mechanism, the photon blockade effect can also be realized 
through alternative physical schemes. These include the quantum Zeno 
effect, which dynamically restricts the system to a decoupled subspace 
\cite{Pelka2025}, a bi-tone drive scheme that induces destructive 
interference without relying on the cavity's inherent nonlinearity 
\cite{Li2022}, and the utilization of two-photon absorption to suppress 
higher-order excitations \cite{Zhou2022}. The UPB has also been extended 
to more complex systems to enrich its physical mechanisms. For example, 
Mahana \textit{et al.} investigated UPB in a cQED system with 
degenerate parametric amplification \cite{Mahana2026}, Sarma \textit{et al.} 
studied the effect in a three-mode optomechanical system \cite{Sarma2018}, 
and Shen \textit{et al.} proposed a scheme for nonreciprocal UPB in a 
driven dissipative cavity with parametric amplification \cite{Shen2020}. 
Furthermore, Xia \textit{et al.} demonstrated a giant nonreciprocal UPB 
using a single atom in an asymmetric Fabry-Pérot cavity, where the optimal 
atom-cavity coupling and the resulting photon statistics depend sensitively 
on the driving direction \cite{Xia2021}, while the influence of 
non-Markovian effects on UPB in driven dissipative coupled cavities was 
analyzed by Shen \textit{et al.} \cite{Shen2024}.

Among various nonlinear platforms, the coupled-cavity system with a 
second-order nonlinearity (\(\chi^{(2)}\)) has attracted particular 
attention, because this nonlinearity can mediate direct frequency 
conversion between different optical modes. For instance, 
Zhou \textit{et al.} \cite{zhou2015} theoretically investigated the UPB 
in a doubly resonant two-cavity system coupled via a second-order nonlinearity. They demonstrated that under simultaneous bimodal driving, 
the destructive quantum interference between different excitation 
pathways can yield strong photon antibunching, even in the 
weak-nonlinearity regime. The optimal condition was derived as 
\(F_B = -F_A^2/\textsl{g}\). However, this analytical derivation relies on two 
idealized assumptions: symmetric cavity dissipation (\(\kappa_A = \kappa_B\)) 
and exact resonance (\(\Delta = 0\)). In realistic experimental 
implementations, such as nanostructured or superconducting cavities, 
fabrication imperfections and material inhomogeneities inevitably lead 
to unequal decay rates, and frequency detuning is usually non-negligible. 
These realistic conditions readily break the delicate \(\pi\) phase 
matching required for the interference, severely degrading the blockade 
effect and thus limiting the practical applicability of the idealized 
condition.

In this paper, we demonstrate that dissipation asymmetry can serve as 
an active and constructive degree of freedom to overcome this limitation. 
We consider a double-cavity system with a \(\chi^{(2)}\) nonlinearity, 
where the dissipation rates of the two cavities are treated as independent 
parameters. By deriving the analytical optimal conditions, we identify 
two solution families. Crucially, the second family, which emerges at a 
specific asymmetric ratio \(\kappa_B = 2\kappa_A\), locks the phase of 
the interference pathways. We reveal that this asymmetric dissipation 
not only protects the optimal blockade condition against arbitrary 
detuning but also significantly deepens the antibunching valley, without 
compromising the output brightness.

The remainder of this paper is organized as follows. In Sec.~\ref{sec:model}, 
we introduce the physical model of the \(\chi^{(2)}\)-nonlinear coupled 
cavities and derive the effective non-Hermitian Hamiltonian. In 
Sec.~\ref{sec:analytical}, we analytically solve the steady-state 
amplitudes and derive the two families of optimal blockade conditions. 
In Sec.~\ref{sec:numerical_method}, we describe the numerical method based 
on the Lindblad master equation. The numerical results and the underlying 
physical mechanism are discussed in Sec.~\ref{sec:results} and Sec.~\ref{sec:mechanism}, respectively. In Sec.~\ref{sec:comparison}, we 
compare our findings with previous works to highlight the robustness of 
the asymmetric dissipation scheme. Finally, we conclude in 
Sec.~\ref{sec:conclusion}.
 
\section{Physical Model}
\label{sec:model}
\begin{figure}[htb]
  \centering
  \includegraphics[width=0.85\columnwidth]{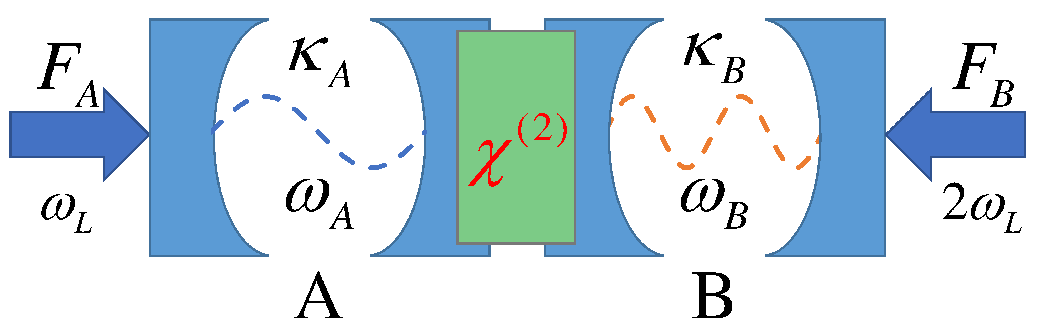}
  \caption{Schematic of the \(\chi^{(2)}\)-nonlinear coupled double-cavity
  system. Cavity A (mode frequency \(\omega_A\)) is driven by a laser field
  of amplitude \(F_A\) and frequency \(\omega_L\), while cavity B (mode
  frequency \(\omega_B\)) is driven by a field of amplitude \(F_B\) and
  frequency \(2\omega_L\). The two cavities have decay rates \(\kappa_A\)
  and \(\kappa_B\), and are coupled via a \(\chi^{(2)}\) nonlinear medium
  (green region) that mediates the conversion between two photons in mode
  A and one photon in mode B.}
  \label{fig:fig1}
\end{figure}

We consider two single-mode cavities, A and B, coupled through a
second-order nonlinearity, as illustrated in Fig.~\ref{fig:fig1}. The two cavities have frequencies
$\omega_A$ and $\omega_B$, respectively, with $\omega_B = 2\omega_A$.
The spatial modes of the two cavities overlap, and the cavities are
coupled via a $\chi^{(2)}$ nonlinearity that converts one photon in
cavity B into two photons in cavity A. The cavities are driven
externally: cavity A by a field of amplitude $F_A$ and frequency
$\omega_L$, and cavity B by a field of amplitude $F_B$ and frequency
$2\omega_L$. The Hamiltonian of the system is written as
\begin{align}
  \hat{H} = {} & \omega_A \hat{a}^\dagger \hat{a}
               + \omega_B \hat{b}^\dagger \hat{b}
               + \textsl{g}\left[\left(\hat{a}^\dagger\right)^2 \hat{b}
               + \hat{b}^\dagger \hat{a}^2\right] \nonumber \\
             & + F_A\left(\hat{a}^\dagger e^{-i\omega_L t}
               + \hat{a}\, e^{i\omega_L t}\right) \nonumber \\
             & + F_B\left(\hat{b}^\dagger e^{-i2\omega_L t}
               + \hat{b}\, e^{i2\omega_L t}\right),
  \label{eq:hamiltonian}
\end{align}
where $\hat{a}^\dagger$ ($\hat{a}$) and $\hat{b}^\dagger$ ($\hat{b}$)
are the creation (annihilation) operators of the modes in cavities A
and B, respectively, and $\textsl{g}$ denotes the second-order nonlinear
coupling strength.

The nonlinear interaction strength \(\textsl{g}\) can be expressed in 
terms of the classical electric-field profiles of the two modes as in Refs. 
\cite{PhysRevB.87.235319,Rodriguez2007}
\begin{equation}
  \textsl{g} = D \epsilon_{0} \left( \frac{\hbar\omega_{A}}{2\epsilon_{0}} \right) 
  \sqrt{\frac{\hbar\omega_{B}}{2\epsilon_{0}}} 
  \int d\mathbf{r}\, 
  \frac{\chi^{(2)}(\mathbf{r})}{\left[\varepsilon_r(\mathbf{r})\right]^{3/2}}\, 
  u_{A}^{2}(\mathbf{r})\,u_{B}(\mathbf{r}),
  \label{eq:g_coefficient}
\end{equation}
where \(\epsilon_0\) is the vacuum permittivity, \(D\) is the 
degeneracy factor accounting for the number of equivalent terms 
in the second-order nonlinear tensor (e.g., \(D=2\) for 
second-harmonic generation), \(\varepsilon_r(\mathbf{r})\) is 
the relative permittivity distribution, \(\chi^{(2)}(\mathbf{r})\) 
is the spatially varying second-order nonlinear susceptibility, 
and \(u_{A,B}(\mathbf{r})\) are the normalized dimensionless 
mode profiles for modes \(A\) and \(B\).

To investigate the system dynamics, we transform to a frame rotating
at the drive frequencies via
$\hat{U}(t)=\exp\left[i\omega_L t\left(\hat{a}^{\dagger}\hat{a}
+2\hat{b}^{\dagger}\hat{b}\right)\right]$, which rotates modes A and B at
$\omega_L$ and $2\omega_L$, respectively. The choice of \(2\omega_L\) for mode B 
follows from the two-photon resonance condition \(\omega_B=2\omega_A\) of the \(\chi^{(2)}\)
process and removes the explicit time dependence of the drives. Using
$\hat{H}_{\mathrm{eff}}=\hat{U}\hat{H}\hat{U}^{\dagger}
-i\hat{U}\,d\hat{U}^{\dagger}/dt$, we obtain
\begin{align}
  \hat{H}_{\mathrm{eff}}
   ={} & \Delta\hat{a}^{\dagger}\hat{a}
      +2\Delta\hat{b}^{\dagger}\hat{b}
      +\textsl{g}\left[\left(\hat{a}^{\dagger}\right)^{2}\hat{b}
       +\hat{a}^{2}\hat{b}^{\dagger}\right] \nonumber \\
     &+F_A\left(\hat{a}^{\dagger}+\hat{a}\right)
      +F_B\left(\hat{b}^{\dagger}+\hat{b}\right),
     \label{eq:Heff}
\end{align}
with $\Delta=\Delta_A=\omega_A-\omega_L$ and
$2\Delta=\Delta_B=\omega_B-2\omega_L$. Because $\omega_B=2\omega_A$,
the detunings satisfy $\Delta_B=2\Delta_A$, reflecting the
two-photon resonance between the drives and the nonlinear process.
The time-independent $\hat{H}_{\mathrm{eff}}$ is used in all
steady-state calculations below.

We now include the dissipation of the two cavities. In the
weak-coupling regime, the quantum-jump terms of the
Lindblad dissipators can be neglected \cite{breuer2002theory, PhysRevLett.68.580, RevModPhys.70.101},
and the conditional dynamics of the system is then governed by the
non-Hermitian Hamiltonian
\begin{equation}
\begin{aligned}
  \tilde{H}
  &=\hat{H}_{\mathrm{eff}}
    -i\frac{\kappa_A}{2}\hat{a}^{\dagger}\hat{a}
    -i\frac{\kappa_B}{2}\hat{b}^{\dagger}\hat{b} \\
  &=\left(\Delta-i\frac{\kappa_A}{2}\right)\hat{a}^{\dagger}\hat{a}
   +\left(2\Delta-i\frac{\kappa_B}{2}\right)\hat{b}^{\dagger}\hat{b}  \\
  &\quad +\textsl{g}\left[\left(\hat{a}^{\dagger}\right)^{2}\hat{b}
     +\hat{a}^{2}\hat{b}^{\dagger}\right]+F_A\left(\hat{a}^{\dagger}+\hat{a}\right)
   +F_B\left(\hat{b}^{\dagger}+\hat{b}\right),
\end{aligned}
\label{eq:Heff_nonherm}
\end{equation}
where \(\kappa_A\) and \(\kappa_B\) are the photon decay rates of
cavities A and B, respectively, and \(\hat{H}_{\mathrm{eff}}\) is
the effective Hamiltonian of Eq.~\eqref{eq:Heff}. The non-Hermitian
terms \(-i\kappa_k\hat{a}_k^{\dagger}\hat{a}_k/2\) (with
\(\hat{a}_A\equiv\hat{a}\) and \(\hat{a}_B\equiv\hat{b}\)) describe
the loss of cavity photons to the external environment and
correspond to the anticommutator part of the standard Lindblad
dissipators,
\begin{equation}
  \mathcal{L}_k[\hat{\rho}]
   =\kappa_k\left[\hat{a}_k\hat{\rho}\hat{a}_k^{\dagger}
    -\frac{1}{2}\left\{ \hat{a}_k^{\dagger}\hat{a}_k, \hat{\rho} \right\}\right],
    \quad k=A,B.
  \label{eq:lindblad}
\end{equation}

The two decay rates \(\kappa_A\) and \(\kappa_B\) are, in general,
independent parameters set by the respective cavity designs. Two
distinct regimes can therefore be identified. When
\(\kappa_A=\kappa_B\), the two cavities are subject to symmetric
dissipation, a situation that is widely assumed in most
previous studies of photon blockade in coupled-cavity systems~\cite{PhysRevA.109.043714,Zhou2026,Xia2026Nonreciprocal}. When \(\kappa_A\neq\kappa_B\), by contrast, the system
operates in the regime of asymmetric dissipation: the two modes
acquire different effective linewidths, so that the competition
between the coherent \(\chi^{(2)}\) conversion and the incoherent
cavity losses is no longer balanced between the two modes. This
dissipation asymmetry constitutes an additional, and so far largely
unexplored, control knob for the blockade mechanism. In the
following, we therefore treat the ratio \(\kappa_B/\kappa_A\) as an
independent parameter of the model, rather than as a fixed property
of the structure.
\section{Analytical Results}
\label{sec:analytical}
To capture the essential physics of the photon blockade, we truncate
the Hilbert space to the low-excitation subspace spanned by the Fock
states \(\lvert 00\rangle\), \(\lvert 10\rangle\), \(\lvert 01\rangle\),
and \(\lvert 20\rangle\). The state of the system is thus expanded as
\begin{equation}
  \lvert\psi\rangle
  =C_{00}\lvert 00\rangle
  +C_{10}\lvert 10\rangle
  +C_{01}\lvert 01\rangle
  +C_{20}\lvert 20\rangle.
  \label{eq:ansatz}
\end{equation}

Substituting this ansatz together with the non-Hermitian Hamiltonian
\(\tilde{H}\) of Eq.~\eqref{eq:Heff_nonherm} into the Schrödinger
equation
\begin{equation}
  i\frac{\partial}{\partial t}\lvert\psi\rangle
  =\tilde{H}\lvert\psi\rangle,
  \label{eq:schrodinger}
\end{equation}
and projecting onto each basis state, we obtain the coupled equations
of motion for the probability amplitudes,
\begin{align}
  i\frac{\partial C_{00}}{\partial t}
  &=F_A C_{10}+F_B C_{01},
  \label{eq:eom00}\\
  i\frac{\partial C_{10}}{\partial t}
  &=F_A C_{00}
   +\left(\Delta-i\frac{\kappa_A}{2}\right)C_{10}
   +\sqrt{2}\,F_A C_{20},
  \label{eq:eom10}\\
  i\frac{\partial C_{01}}{\partial t}
  &=F_B C_{00}
   +\left(2\Delta-i\frac{\kappa_B}{2}\right)C_{01}
   +\sqrt{2}\,\textsl{g}\,C_{20},
  \label{eq:eom01}\\
  i\frac{\partial C_{20}}{\partial t}
  &=\sqrt{2}\,F_A C_{10}
   +\sqrt{2}\,\textsl{g}\,C_{01}
   +2\left(\Delta-i\frac{\kappa_A}{2}\right)C_{20}.
  \label{eq:eom20}
\end{align}

Here the factors of \(\sqrt{2}\) originate from the bosonic
raising and lowering operators, e.g.,
\(\hat{a}^{\dagger}\lvert 10\rangle=\sqrt{2}\lvert 20\rangle\) and
\(\hat{a}\lvert 20\rangle=\sqrt{2}\lvert 10\rangle\), and the
complex coefficients \(\Delta-i\kappa_A/2\) and
\(2\Delta-i\kappa_B/2\) encode the detunings and decay rates of the
two cavity modes introduced in Eq.~\eqref{eq:Heff_nonherm}.

In the steady state, the time derivatives in
Eqs.~\eqref{eq:eom00}--\eqref{eq:eom20} vanish, which yields the
algebraic system
\begin{align}
  0&=F_A C_{10}+F_B C_{01},
  \label{eq:steady00}\\
  0&=F_A C_{00}
   +\left(\Delta-i\frac{\kappa_A}{2}\right)C_{10}
   +\sqrt{2}\,F_A C_{20},
  \label{eq:steady10}\\
  0&=\sqrt{2}\,F_A C_{10}
   +2\left(\Delta-i\frac{\kappa_A}{2}\right)C_{20}
   +\sqrt{2}\,\textsl{g}\,C_{01},
  \label{eq:steady20}\\
  0&=F_B C_{00}
   +\sqrt{2}\,\textsl{g}\,C_{20}
   +\left(2\Delta-i\frac{\kappa_B}{2}\right)C_{01}.
  \label{eq:steady01}
\end{align}

Under the weak-driving condition \(F_A, F_B \ll \kappa_A, \kappa_B\),
the ground-state amplitude overwhelmingly dominates, establishing the
hierarchy \(C_{00} \gg C_{10} \gg \{C_{01}, C_{20}\}\). Consequently, the
normalization condition 
\(|C_{00}|^{2} + |C_{10}|^{2} + |C_{01}|^{2} + |C_{20}|^{2} = 1\) allows
us to safely set \(C_{00} \simeq 1\). Within this
perturbative framework, the equation for the vacuum state,
Eq.~\eqref{eq:steady00}, only governs the higher-order balance and is
decoupled from the leading-order dynamics of the excited states.
Therefore, it is omitted when determining the amplitudes. The
steady-state solution is then given by retaining the remaining three
coupled equations (with \(C_{00}=1\)):
\begin{align}
  0&=F_A
   +\left(\Delta-i\frac{\kappa_A}{2}\right)C_{10}
   +\sqrt{2}\,F_A C_{20},
  \label{eq:steady10s}\\
  0&=\sqrt{2}\,F_A C_{10}
   +2\left(\Delta-i\frac{\kappa_A}{2}\right)C_{20}
   +\sqrt{2}\,\textsl{g}\,C_{01},
  \label{eq:steady20s}\\
  0&=F_B
   +\sqrt{2}\,\textsl{g}\,C_{20}
   +\left(2\Delta-i\frac{\kappa_B}{2}\right)C_{01}.
  \label{eq:steady01s}
\end{align}

For the steady state, solving Eqs.~\eqref{eq:steady10s}--
\eqref{eq:steady01s} symbolically (e.g., with Wolfram Mathematica)
yields closed-form expressions for the probability amplitudes,
\begin{align}
  C_{10}&=-\frac{F_A\left(D_AD_B-\textsl{g}^2+\textsl{g}F_B\right)}{N},
  \label{eq:C10exact}\\
  C_{01}&=\frac{F_A^2(F_B-\textsl{g})-F_BD_A^2}{N},
  \label{eq:C01exact}\\
  C_{20}&=\frac{F_A^2D_B+\textsl{g}F_BD_A}{\sqrt{2}\,N},
  \label{eq:C20exact}
\end{align}
where
\(D_A\equiv\Delta-i\kappa_A/2\) and
\(D_B\equiv2\Delta-i\kappa_B/2\)
are the complex detunings of modes A and B, respectively, and
\(N\equiv D_A^2D_B-F_A^2D_B-\textsl{g}^2D_A\) is the common denominator
of the three amplitudes.

Photon blockade in cavity A requires the two-photon amplitude to
vanish, \(C_{20}=0\). Since the denominator of \(C_{20}\) is
nonvanishing in the weak-driving regime, this condition is
equivalent to requiring its numerator to be zero, i.e., 
\(F_A^2 D_B + \textsl{g} F_B D_A = 0\). Substituting 
\(D_A = \Delta - i\kappa_A/2\) and \(D_B = 2\Delta - i\kappa_B/2\) 
into this condition, we obtain
\begin{equation}
  4\Delta F_A^2+2\Delta \textsl{g}F_B-i\textsl{g}F_B\kappa_A-iF_A^2\kappa_B=0.
  \label{eq:C20_zero}
\end{equation}

Because all parameters \(F_A\), \(F_B\), \(\textsl{g}\), \(\kappa_A\),
\(\kappa_B\), and \(\Delta\) are real, the real and imaginary parts
of Eq.~\eqref{eq:C20_zero} must vanish independently,
\begin{align}
  4\Delta F_A^2+2\Delta \textsl{g}F_B&=0,
  \label{eq:real_part}\\
  -\textsl{g}F_B\kappa_A-F_A^2\kappa_B&=0.
  \label{eq:imag_part}
\end{align}

Equation~\eqref{eq:imag_part} yields
\begin{equation}
  F_B=-\frac{F_A^2\kappa_B}{\textsl{g}\kappa_A},
  \label{eq:FB_imag}
\end{equation}
and substituting Eq.~\eqref{eq:FB_imag} into
Eq.~\eqref{eq:real_part} gives
\begin{equation}
  \Delta\left(2-\frac{\kappa_B}{\kappa_A}\right)F_A^2=0.
  \label{eq:Delta_factor}
\end{equation}

For nonzero drives (\(F_A\neq0\)) and nonvanishing coupling
(\(\textsl{g}\neq0\)), Eq.~\eqref{eq:Delta_factor} requires either
\(\Delta=0\) or \(\kappa_B=2\kappa_A\). Substituting each branch
back into Eq.~\eqref{eq:FB_imag} then determines the corresponding
drive amplitude, and two solution families emerge:
\begin{align}
  \text{Family 1:}\quad
  &\Delta=0,\quad
   F_B=-\frac{F_A^2\kappa_B}{\textsl{g}\kappa_A},
  \label{eq:family1}\\[4pt]
  \text{Family 2:}\quad
  &\kappa_B=2\kappa_A,\quad
   F_B=-\frac{2F_A^2}{\textsl{g}},
  \label{eq:family2}
\end{align}
where Family~1 holds for arbitrary \(\kappa_A\) and \(\kappa_B\),
and Family~2 holds for arbitrary \(\Delta\). These two families
constitute the optimal conditions for photon blockade in cavity A.
Family~1 requires resonant tuning (\(\Delta=0\)) between the cavity
mode and the drive, whereas Family~2 is protected by the dissipation
asymmetry \(\kappa_B=2\kappa_A\) and thus persists for arbitrary
detuning, which makes Family 2 significantly more robust against detuning.

Before turning to the detailed numerical results, we note that the
dissipation asymmetry introduced above is not a passive property of
the structure but an active control parameter. As shown in
Eq.~\eqref{eq:family2}, the optimal blockade condition is not
reached in the symmetric limit \(\kappa_A=\kappa_B\), but rather at
a specific asymmetric ratio, \(\kappa_B=2\kappa_A\). This result
distinguishes the present mechanism from previous proposals, in
which dissipation asymmetry was either absent or treated as a fixed,
non-tunable property of the structure.
\section{Numerical method}
\label{sec:numerical_method}
To numerically investigate the photon blockade effect in the presence
of cavity dissipation, the system dynamics is governed by the Lindblad
master equation,
\begin{align}
  \frac{\partial \hat{\rho}}{\partial t} = -i\left[ \hat{H}_{\mathrm{eff}}, \hat{\rho} \right]
  &+ \frac{\kappa_A}{2}\left( 2\hat{a}\hat{\rho}\hat{a}^{\dagger} - \hat{a}^{\dagger}\hat{a}\hat{\rho} - \hat{\rho}\hat{a}^{\dagger}\hat{a} \right) \nonumber \\
  &+ \frac{\kappa_B}{2}\left( 2\hat{b}\hat{\rho}\hat{b}^{\dagger} - \hat{b}^{\dagger}\hat{b}\hat{\rho} - \hat{\rho}\hat{b}^{\dagger}\hat{b} \right),
  \label{eq:master}
\end{align}
where \(\hat{\rho}\) is the density matrix of the system.
To characterize the steady-state statistical properties of the photons
in cavity A, we evaluate the equal-time second-order correlation
function, defined as
\begin{equation}
  g^{(2)}(0) = \frac{\langle \hat{a}^{\dagger}\hat{a}^{\dagger}\hat{a}\hat{a} \rangle_{\mathrm{ss}}}{\langle \hat{a}^{\dagger}\hat{a} \rangle_{\mathrm{ss}}^2}
  = \frac{\mathrm{Tr}\left( \hat{\rho}_{\mathrm{ss}} \hat{a}^{\dagger}\hat{a}^{\dagger}\hat{a}\hat{a} \right)}{\left[ \mathrm{Tr}\left( \hat{\rho}_{\mathrm{ss}} \hat{a}^{\dagger}\hat{a} \right) \right]^2},
  \label{eq:g2_numerical}
\end{equation}
where \(\langle \cdot \rangle_{\mathrm{ss}}\) denotes the steady-state expectation value, \(\mathrm{Tr}(\cdot)\) denotes the trace operation, and \(\hat{\rho}_{\mathrm{ss}}\) is the steady-state density matrix.
For a single-photon source, the brightness is a key figure of merit
that directly reflects the capability of single-photon emission.
Conventionally, the brightness is quantified by the mean 
photon number,
\begin{equation}
  N_a = \langle \hat{a}^{\dagger}\hat{a} \rangle_{\mathrm{ss}} = \mathrm{Tr}\left( \hat{\rho}_{\mathrm{ss}} \hat{a}^{\dagger}\hat{a} \right).
  \label{eq:mean_photon}
\end{equation}

Given the effective Hamiltonian of the system, the steady-state density matrix \(\hat{\rho}_{\mathrm{ss}}\) is obtained by numerically solving Eq.~\eqref{eq:master}. Substituting \(\hat{\rho}_{\mathrm{ss}}\) into Eqs.~\eqref{eq:g2_numerical} and \eqref{eq:mean_photon} yields the numerical solutions for \(g^{(2)}(0)\) and \(N_a\), respectively, thereby allowing us to quantitatively investigate the photon blockade effect.

All numerical simulations presented in this work are carried out using the open-source Quantum Optics Toolbox for MATLAB \cite{Tan_1999}, combined with our custom-written codes for the computations and data plotting. To facilitate a unified numerical treatment, the decay rate of cavity A, \(\kappa_A\), is chosen as the fundamental scale, and all other physical quantities are rendered dimensionless in units of \(\kappa_A\).

\section{Results and Discussion}
\label{sec:results}
\subsection{Enhancement of photon blockade via asymmetric dissipation}

\begin{figure}[htb]
  \centering
  \includegraphics[width=\columnwidth]{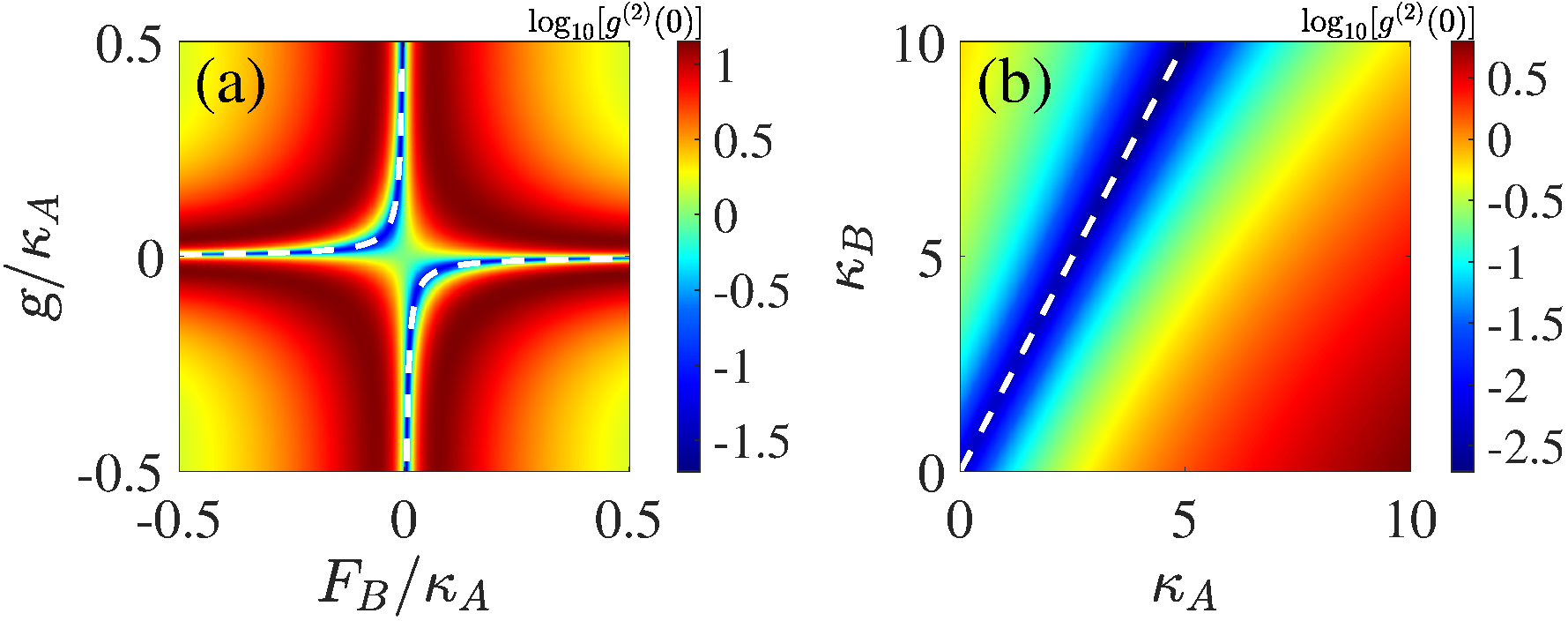}
  \caption{Density plots of the logarithmic equal-time second-order correlation 
  function \(\log_{10}[g^{(2)}(0)]\) in different 
  parameter spaces. (a) The dependence on the B-cavity drive strength 
  \(F_B\) and the nonlinear coupling \(\textsl{g}\), with fixed parameters 
  \(\Delta=0\), \(\kappa_A=\kappa_B=1.0\), and \(F_A=0.05\). The white 
  dashed line indicates the analytical optimal condition of Family~1: 
  \(F_B = -F_A^2\kappa_B/(\textsl{g}\kappa_A)\). The minimum value in this map 
  reaches approximately \(g^{(2)}(0) \sim 10^{-1.5}\). (b) The dependence 
  on the cavity decay rates \(\kappa_A\) and \(\kappa_B\), with fixed 
  parameters \(\Delta=2.0\) , \(\textsl{g}=0.15\), \(F_A=0.05\), 
  and \(F_B = -2F_A^2/\textsl{g}= - 0.0333\). The white dashed line indicates the analytical 
  optimal condition of Family~2: \(\kappa_B = 2\kappa_A\). Notably, 
  even with a finite detuning, the minimum value here approaches 
  \(g^{(2)}(0) \sim 10^{-2.5}\), which is an order of magnitude lower 
  than that in (a). This demonstrates that asymmetric dissipation 
  significantly enhances the photon blockade and protects it against 
  detuning.}
  \label{fig:fig2}
\end{figure}

To verify our analytical predictions and to comprehensively explore the parameter space for the UPB, we perform numerical simulations of the master equation in a truncated Fock space. The results for the equal-time second-order correlation function \(g^{(2)}(0)\) are presented in Fig.~\ref{fig:fig2}, where the white dashed lines represent our analytical optimal conditions derived in Eqs.~\eqref{eq:family1} and \eqref{eq:family2}.

Figure~\hyperref[fig:fig2]{\ref*{fig:fig2}(a)} displays the logarithmic map of \(\log_{10}[g^{(2)}(0)]\) as a function of the B-cavity drive strength \(F_B\) and the nonlinear coupling \(\textsl{g}\), with the fixed parameters given in the caption. The strong photon antibunching region exhibits a clear hyperbolic dependence, which perfectly aligns with the analytical prediction. This hyperbolic behavior implies that the optimal interference condition \(F_B = -F_A^2\kappa_B/(\textsl{g}\kappa_A)\) requires the drive \(F_B\) and the coupling \(\textsl{g}\) to have opposite signs. This sign relationship ensures the essential \(\pi\) phase difference between the direct two-photon transition pathway and the cascade pathway, leading to complete destructive interference. The minimum value in this map reaches approximately \(g^{(2)}(0) \sim 10^{-1.5}\). When the parameters deviate from this optimal valley, the quantum interference is broken, causing the correlation function to increase 
rapidly. In the background regions far from the optimal condition, 
\(g^{(2)}(0)\) exceeds unity (i.e., \(\log_{10}[g^{(2)}(0)] > 0\)), 
indicating a transition into the photon-bunching regime where the 
blockade effect is completely lost.

Next, we investigate the role of dissipation asymmetry by plotting \(\log_{10}[g^{(2)}(0)]\) in the \(\kappa_A\)-\(\kappa_B\) parameter plane, as shown in Fig.~\hyperref[fig:fig2]{\ref*{fig:fig2}(b)}. Remarkably, even in the presence of a finite detuning \(\Delta=2.0\), the numerical deep-blue valley forms a straight line that strictly follows \(\kappa_B = 2\kappa_A\), in excellent agreement with the condition of Family~2. Notably, the suppression of two-photon excitation here is even more pronounced: the minimum value approaches \(g^{(2)}(0) \sim 10^{-2.5}\), which is an order of magnitude lower than that in Fig.~\hyperref[fig:fig2]{\ref*{fig:fig2}(a)}. This significant reduction demonstrates that the asymmetric dissipation not only protects the UPB condition against detuning but also enhances the blockade effect itself. As the system deviates from this optimal ratio towards the symmetric limit \(\kappa_A = \kappa_B\), the value of \(g^{(2)}(0)\) increases drastically, and the antibunching effect deteriorates. This feature clearly demonstrates that the dissipation asymmetry acts as an active and constructive degree of freedom for optimizing single-photon sources.

\subsection{Detuning dependence and brightness comparison}
\begin{figure}[htb]
  \centering
  \includegraphics[width=\columnwidth]{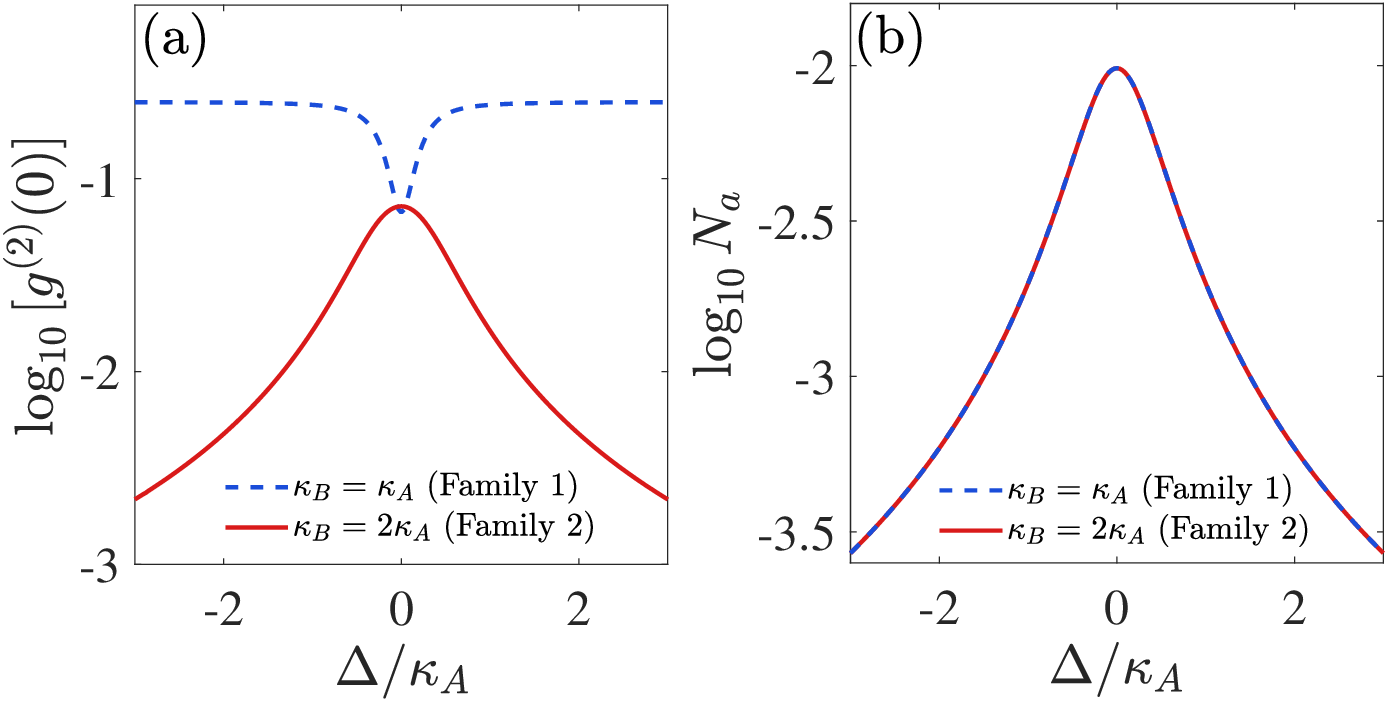}
  \caption{Comparison of the two solution families as a 
  function of the normalized detuning \(\Delta/\kappa_A\). The common 
  fixed parameters are \(\kappa_A = 1.0\), \(F_A = 0.05\), and \(\textsl{g} = 0.15\). 
  The dashed blue line corresponds to Family~1, where 
  \(\kappa_B = \kappa_A = 1.0\) and \(F_B = -F_A^2\kappa_B/(\textsl{g}\kappa_A) \approx -0.0167\). 
  The solid red line corresponds to Family~2, where 
  \(\kappa_B = 2\kappa_A = 2.0\) and \(F_B = -2F_A^2/\textsl{g} \approx -0.0333\). 
  (a) The logarithmic equal-time second-order correlation function 
  \(\log_{10}[g^{(2)}(0)]\). The blockade for Family~1 is highly sensitive 
  to detuning and only survives near \(\Delta=0\), whereas Family~2 
  maintains a deep and flat antibunching valley across the broad off-resonant 
  region, demonstrating strong robustness against detuning. 
  (b) The logarithmic mean photon number \(\log_{10} N_a\). 
  The two curves almost overlap, indicating that the asymmetric 
  dissipation enhances the blockade without sacrificing the brightness 
  of the single-photon source.}
  \label{fig:fig3}
\end{figure}

To further elucidate the profound differences between the two solution 
families, we plot the logarithmic equal-time second-order correlation
function \(\log_{10}[g^{(2)}(0)]\) and the mean photon number \(\log_{10} N_a\)  
as functions of the normalized detuning \(\Delta/\kappa_A\) in 
Fig.~\ref{fig:fig3}. The dashed blue line corresponds to Family~1 
(\(\kappa_A=\kappa_B\)), while the solid red line represents Family~2 
(\(\kappa_B=2\kappa_A\)).

As illustrated in Fig.~\hyperref[fig:fig3]{\ref*{fig:fig3}(a)}, the two families exhibit 
drastically different behaviors against the detuning. For Family~1, the 
photon blockade is strictly restricted to the exact resonance 
(\(\Delta=0\)), where \(g^{(2)}(0)\) reaches a local minimum of 
approximately \(10^{-1.1}\). Any finite detuning causes the blockade 
effect to deteriorate rapidly. In stark contrast, Family~2 maintains an 
extremely deep antibunching effect over the entire scanning range of 
\(\Delta/\kappa_A \in [-3, 3]\), with \(g^{(2)}(0)\) stabilizing around 
\(10^{-2.7}\). This flat and deep valley firmly demonstrates that the 
asymmetric dissipation \(\kappa_B = 2\kappa_A\) renders the optimal 
blockade condition completely immune to frequency detuning.

Remarkably, as shown in Fig.~\hyperref[fig:fig3]{\ref*{fig:fig3}(b)}, the mean photon 
numbers for both solution families almost coincide perfectly. This 
implies that the implementation of asymmetric dissipation does not 
compromise the brightness of the single-photon source. Therefore, 
compared to Family~1, Family~2 provides a superior scheme that 
simultaneously achieves a much deeper photon blockade and extreme 
robustness against detuning, without sacrificing the output intensity.

\section{Physical Mechanism of Unconventional Photon Blockade}
\label{sec:mechanism}
\begin{figure}[htb]
  \centering
  \includegraphics[width=0.7\columnwidth]{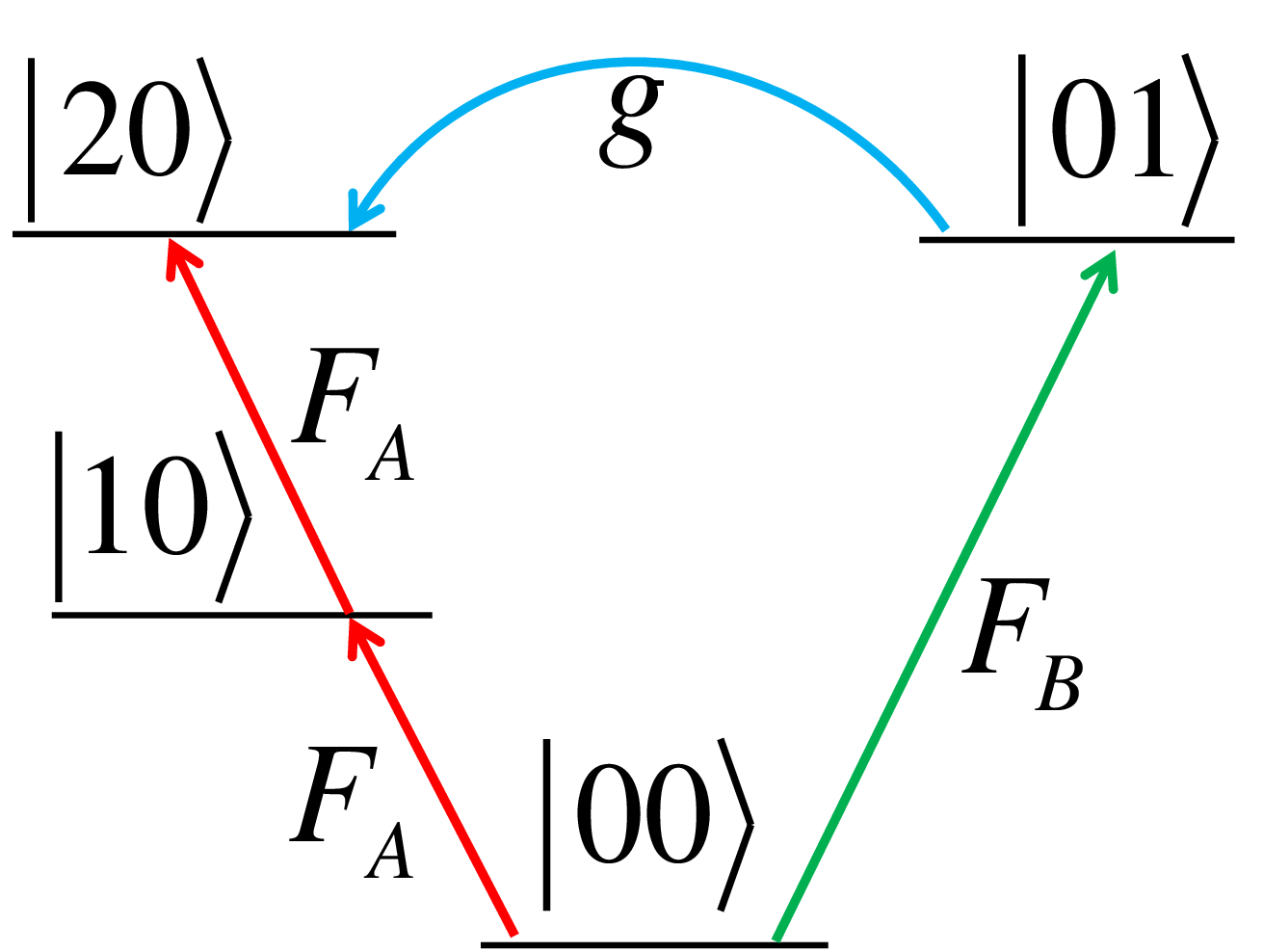}
  \caption{Energy-level diagram illustrating the quantum interference 
  mechanism responsible for photon blockade in cavity A. The system is 
  restricted to the subspace spanned by the Fock states \(|00\rangle\), 
  \(|10\rangle\), \(|01\rangle\), and \(|20\rangle\). Two distinct pathways 
  lead to the two-photon state \(|20\rangle\): the direct pathway via two 
  consecutive excitations by the external drive \(F_A\) on cavity A (red 
  arrows), and the cascade pathway where the external drive \(F_B\) first 
  excites cavity B (green arrow) followed by the \(\chi^{(2)}\) nonlinear 
  conversion (blue arrow, strength \(\textsl{g}\)). The destructive interference 
  between these two pathways gives rise to the unconventional photon blockade.}
  \label{fig:energy_level}
\end{figure}

To explore the physical mechanism of the UPB in the system, we illustrate the possible photon transition pathways from the ground state to the two-photon state in Fig.~\ref{fig:energy_level}. As shown in the energy-level diagram, there exist two distinct pathways for the system to reach the two-photon state \(|20\rangle\). The first pathway is the direct two-photon excitation within cavity A, where the system is driven consecutively by the external field \(F_A\) to transition \(|00\rangle \xrightarrow{F_A} |10\rangle \xrightarrow{F_A} |20\rangle\). The second pathway is a cascade process involving both cavities: the external drive \(F_B\) first excites cavity B to the single-photon 
state \(|01\rangle\), which then transitions to the two-photon state of cavity A via the \(\chi^{(2)}\) nonlinear interaction \(\textsl{g}\), 
i.e., \(|00\rangle \xrightarrow{F_B} |01\rangle \xrightarrow{\textsl{g}} |20\rangle\).

When the system parameters satisfy the optimal conditions given by 
either Eq.~\eqref{eq:family1} or Eq.~\eqref{eq:family2}, the two 
pathways interfere destructively, completely canceling the probability 
amplitude of the two-photon state \(|20\rangle\). Consequently, the 
system predominantly remains in the single-photon or vacuum states, 
exhibiting the photon antibunching statistics characteristic of a 
single-photon source. In stark contrast to the CPB, which relies on strong single-photon nonlinearities to 
induce anharmonic energy-level splitting, the physical origin of the 
blockade in our scheme is purely multipath quantum interference. 
Therefore, the UPB mechanism relaxes the stringent requirement for 
strong nonlinearity, significantly facilitating its experimental 
realization.

\section{Comparison With Previous Works And Robustness Against Detuning}
\label{sec:comparison}
It is instructive to compare the solution families derived above with the 
results of previous works, specifically Ref.~\cite{zhou2015}, as summarized 
in Table~\ref{tab:conditions}. In Ref.~\cite{zhou2015}, the analysis is 
restricted to the symmetric-dissipation limit \(\kappa_A = \kappa_B = \kappa\) 
with zero detuning \(\Delta = 0\), yielding the optimal condition 
\(F_B = -F_A^2/\textsl{g}\). This previous result is perfectly reproduced 
by taking the symmetric limit \(\kappa_A = \kappa_B\) in our Family~1. 
While Family~1 extends the result of Ref.~\cite{zhou2015} to arbitrary 
dissipation rates, it still strictly requires resonant tuning \(\Delta = 0\); 
any finite detuning will break the delicate destructive interference and 
destroy the blockade.

\begin{table*}[htb]
\centering
\caption{Comparison of the optimal conditions and robustness for different dissipation regimes. The result of Ref.~\cite{zhou2015} is a symmetric special case of Family~1. Family~2, driven by asymmetric dissipation, represents a novel mechanism immune to detuning.}
\label{tab:conditions}
\renewcommand{\arraystretch}{1.5} 
\begin{ruledtabular}
\begin{tabular}{l c c l}
\textbf{Case} & \textbf{Parameters} & \textbf{Optimal condition} & \textbf{Mechanism and robustness} \\
\hline
Ref.~\cite{zhou2015} & 
\(\kappa_A=\kappa_B, \ \Delta=0\) & 
\(F_B = -F_A^2/\textsl{g}\) & 
Symmetric dissipation; sensitive to detuning \\
Family 1 & 
Arbitrary \(\kappa_A, \kappa_B\), \(\Delta=0\) & 
\(F_B = -F_A^2\kappa_B/(\textsl{g}\kappa_A)\) & 
Generalized interference; breaks at \(\Delta \neq 0\) \\
Family 2 & 
\(\kappa_B=2\kappa_A\), any \(\Delta\) & 
\(F_B = -2F_A^2/\textsl{g}\) & 
Phase-locked interference; \textbf{robust against any \(\Delta\)} \\
\end{tabular}
\end{ruledtabular}
\end{table*}

Family~2 represents an entirely new optimal condition that is absent in 
Ref.~\cite{zhou2015}. This novelty originates from a fundamental interplay 
between the nonlinear coupling and the dissipation asymmetry. From the 
interference condition \(F_A^2 D_B + \textsl{g} F_B D_A = 0\), the direct 
two-photon pathway has a complex amplitude proportional to \(F_A^2 D_B\), 
while the cascade pathway is proportional to \(\textsl{g} F_B D_A\). 
By requiring the specific asymmetric ratio \(\kappa_B = 2\kappa_A\), the 
complex detunings automatically satisfy the relation \(D_B = 2D_A\) for 
arbitrary detuning \(\Delta\). Consequently, the condition simplifies to 
\(D_A(2F_A^2 + \textsl{g} F_B) = 0\), which directly yields the real-valued 
blockade condition \(F_B = -2F_A^2/\textsl{g}\). 

To understand the physical mechanism behind this robustness, we examine the 
coherent sum of these complex amplitudes. Destructive interference requires 
the different transition pathways to have equal magnitudes but opposite 
directions in the complex plane, acquiring a relative phase difference of 
\(\pi\). In typical UPB systems, this relative phase is highly sensitive 
to frequency detuning because the pathways accumulate different 
detuning-dependent phases. In Family~2, however, the asymmetric dissipation 
cancels this relative phase accumulation. The detuning factor \(D_A\) merely 
contributes an overall global phase to the system, leaving the ratio of the 
two pathway amplitudes entirely real and independent of \(\Delta\). Therefore, 
the drive condition \(F_B = -2F_A^2/\textsl{g}\) provides the necessary \(\pi\) 
phase difference and keeps the amplitudes perfectly anti-aligned regardless 
of the detuning. In this scenario, the dissipation asymmetry effectively 
locks the relative phase of the two-photon interference pathways, rendering 
the blockade completely immune to variations in \(\Delta\).

This comparison highlights a counterintuitive physical insight: while 
dissipation asymmetry is traditionally viewed as a detrimental factor that 
degrades interference visibility, here it acts as an essential degree of 
freedom that qualitatively reshapes the physical mechanism. From an 
experimental perspective, Family~2 provides a highly robust scheme for 
unconventional photon blockade, as it significantly relaxes the stringent 
requirement of exact resonance, making the phenomenon much more tolerant 
to frequency fluctuations and fabrication imperfections.

\section{Conclusion}
\label{sec:conclusion}
In conclusion, we have theoretically investigated the unconventional photon 
blockade in a second-order-nonlinearity coupled double-cavity system, with 
a particular focus on the role of asymmetric dissipation. By solving the 
non-Hermitian Hamiltonian in the weak-driving limit, we analytically derived 
two families of optimal blockade conditions. Our results show that while 
Family~1 recovers the previous symmetric-dissipation results and requires 
strict resonance \(\Delta=0\), Family~2 arises from a specific asymmetric 
dissipation ratio \(\kappa_B = 2\kappa_A\) and holds for arbitrary detuning.

Through numerical simulations of the Lindblad master equation, we verified 
our analytical predictions. Remarkably, we found that the asymmetric 
dissipation not only protects the unconventional photon blockade effect 
against frequency detuning but also enhances the blockade depth by an order 
of magnitude. Furthermore, the mean photon number remains almost 
identical to that of the symmetric case, indicating that the brightness of 
the single-photon source is preserved. The underlying physical mechanism is 
traced back to the perfect cancellation of the probability amplitude for 
the two-photon state, where the dissipation asymmetry locks 
the relative phase between the direct and cascade transition pathways. Our 
scheme highlights the profound impact of dissipation asymmetry and provides 
a robust and practical route for the design of passive, high-performance 
single-photon sources.

\begin{acknowledgments}
This work is supported by the National Natural Science Foundation of China (NSFC) under Grant No.~12374333, the Jiangxi Provincial Natural Science Foundation under Grant No.~20252BAC200156, and the Early-Career Young Scientists and Technologists Project of Jiangxi Province under Grant No.~20244BCE52197.
\end{acknowledgments}

\section*{DATA AVAILABILITY}
The numerical simulation codes that support the findings of this study are openly available in the GitHub repository at \url{https://github.com/hotbbs/UPB-asymmetric-dissipation/}.

\section*{Conflicts of Interest}
The authors declare no competing interests.

\bibliography{ref.bib}

\end{document}